\documentclass[a4paper,nofootinbib,preprintnumbers,floatfix,superscriptaddress,pra,twocolumn,showpacs]{revtex4-1}
\usepackage{amsmath, amsthm, amssymb}
\usepackage{graphicx}
\usepackage{dcolumn}
\usepackage{bm}
\usepackage{color}
\usepackage{amsmath}
\usepackage{amsfonts}
\usepackage{amssymb}
\usepackage{bbm}
\usepackage{hyperref}

\newcommand{\beq}{\begin{equation}}
\newcommand{\eeq}{\end{equation}}
\newcommand{\bea}[1]{\begin{equation}\begin{array}{#1}}
\newcommand{\eea}{\end{array}\end{equation}}
\newcommand{\beqn}{\begin{eqnarray}}
\newcommand{\eeqn}{\end{eqnarray}}

\renewcommand{\rho}{\varrho}

\newcommand{\processlist}[3][\relax]{%
  \def\listfinish{#1}%
  \long\def\listact{#2}%
  \processnext#3\listfinish}
\newcommand{\processnext}[1]{%
  \ifx\listfinish#1\empty\else\listact{#1}\expandafter\processnext\fi}

\renewcommand{\H}[1]{H(\processlist{X_}{#1})}

\begin{document}
\title{Entropic inequalities as a necessary and sufficient condition to noncontextuality and locality}
\author{Rafael Chaves}
\affiliation{Institute for Physics, University of Freiburg, Rheinstrasse 10, D-79104 Freiburg, Germany}

\begin{abstract}
The assumption of local realism in a Bell locality scenario imposes non-trivial conditions on the Shannon entropies of the associated probability distributions, expressed by linear entropic Bell inequalities. In principle, these entropic inequalities provide necessary but not sufficient criteria for the existence of a local hidden variable model reproducing the correlations, as, for example, the paradigmatic nonlocal PR-box is entropically not different from a classically correlated box. In this paper we show that for the n-cycle scenario, entropic inequalities completely characterize the set of local correlations. In particular, every nonsignalling box which violates the CHSH inequality -- including the PR-box -- can be locally modified so that it also violates the entropic version of CHSH inequality. As we show, any nonlocal probabilistic model when appropriately mixed with a local model, violates an entropic inequality, thus evidencing a very peculiar kind of nonlocality. As the n-cycle captures equally well both the notion of local realism introduced by Bell and that of noncontextuality presented by the Kochen-Specker theorem, the results are also valid for noncontextuality scenarios.
\end{abstract}

 \maketitle
\section{Introduction}
\label{introduction}The quantum nonlocal correlations that may arise in experiments performed by spacelike separated and independent observers, are a key concept in the foundational aspects of quantum mechanics. The expected classical intuition that physical quantities have well-established values previous to any measurement and that
signals cannot propagate instantaneously, do not suffice to reproduce the quantum mechanical predictions~\cite{Bell1964}, highlighting a very counterintuitive aspect of quantum theory that has received strong experimental corroboration over the years~\cite{Freedman1972,*Aspect1982,*Rowe2001,*Matsukevich2008}. From an applied point of view, nonlocality is now recognized as a novel physical resource, which enables protocols such as device-independent quantum key distribution~\cite{Pironio2009}, random number generation~\cite{Pironio2010} and the reduction of communication complexity in distributed-computing scenarios~\cite{Brukner2004,*Buhrman2010}.

In practice, nonlocality is witnessed through the violation of a Bell inequality~\cite{Bell1964}. Given a certain experimental scenario defined by the number of spatially separated parties, the possible different measurement settings for each party, and the possible outcomes for each setting, local-realistic joint probability distributions form a convex set to which Bell inequalities, a set of linear inequalities of the probabilities, are the non-trivial facets \cite{Pitowsky}. This geometric approach provides a general framework in which Bell inequalities can be derived, since the task to find the facets of a convex set is a linear program that can be solved efficiently. The problem is that, generally, the size of the linear program grows very fast as the nonlocality scenario becomes less simple, some classes even being known to be a NP-complete problem~\cite{Pitowsky}. In spite of that, some particular characterizations are well known. The bipartite scenario with two dichotomic measurements per party is completely characterized by the Clauser-Horne-Shimony-Holt (CHSH) inequality~\cite{Clauser1969} and a generalization of CHSH to more outcomes is provided by the Collins-Gisin-Linden-Massar-Popescu (CGLMP) inequality~\cite{Collins2002}, the CHSH and CGLMP inequalities fully describing the set of local correlations up to 3 outcomes~\cite{Masanes2002}. However, for a number of outcomes larger than 3, a complete characterization of the inequalities bounding the set of local correlations is still to be found~\cite{Masanes2002,Bancal2010}, highlighting the difficulty and limitations of this approach.

In a conceptually different approach introduced by Braunstein and Caves~\cite{Braunstein1988}, it was shown that local realism imposes non-trivial conditions already on the level of the Shannon entropies. The Shannon entropies carried by the measurements on two distant systems must satisfy certain inequalities, which can be regarded as entropic Bell inequalities. It was recently pointed out that the characterization of the local correlations on the entropic level, also defines a linear programming problem~\cite{Fritz2012, Chaves2012}. One advantage of this entropic approach is that it can readily be applied to quantum systems of arbitrary local dimension and general measurement operators, since the inequalities do not depend on the number of outcomes of the measured observables. That is, while the dimension of the set of local correlations in terms of probabilities grows exponentially with the number of outcomes for each observable, the entropic set of local correlations is independent of that. Another advantage of the entropic approach is that it easily adapts to situations of additional independence requirements, like the bilocality scenarios introduced by a entanglement swapping experiment~\cite{Zukowski1993, Branciard2010} and general correlation and causal model scenarios~\cite{Pearlbook,Fritz2012_2,Spekkens2012}. The independence constraints are nonlinear on the level of probabilities, defining a nonconvex set, while in terms of entropies such constraints are linear and still define a convex set that can be solved by linear programming. In spite of their attractive properties, entropic Bell inequalities are, in principle, sufficient but not necessary conditions to witness nonlocality. That is, there are nonlocal distributions violating a Bell inequality, that, however, do not violate its entropic counterpart~\cite{Chaves2012}. However, as we show in this paper, the situation is more involved than initially thought, as entropic inequalities can, at least in some scenarios, completely characterize the set of local correlations.

In this paper we show that in the n-cycle scenario, any nonlocal distribution when augmented with shared randomness will also violate a entropic Bell inequality. The n-cycle can be seen as generalization of the CHSH scenario for an arbitrary number of observables for each party, with only a subset of pairwise observables being jointly measurable (Fig.~\ref{fig:context}). The n-cycle captures equally well both the notion of local realism introduced by Bell \cite{Bell1964} and that of noncontextuality presented by the Kochen-Specker (KS) theorem \cite{Kochen1967}. A complete characterization of the n-cycle in terms of an exponential number of tight, linear inequalities has recently been found \cite{Araujo2012}. As we show here, the n-cycle can be equivalently described by a polynomial number of entropic inequalities and a list of local/noncontextual points lying in the facets of the corresponding set of correlations.

\section{The n-cycle scenario}
\begin{figure}[t!]
\begin{center}
\includegraphics[width=\linewidth]{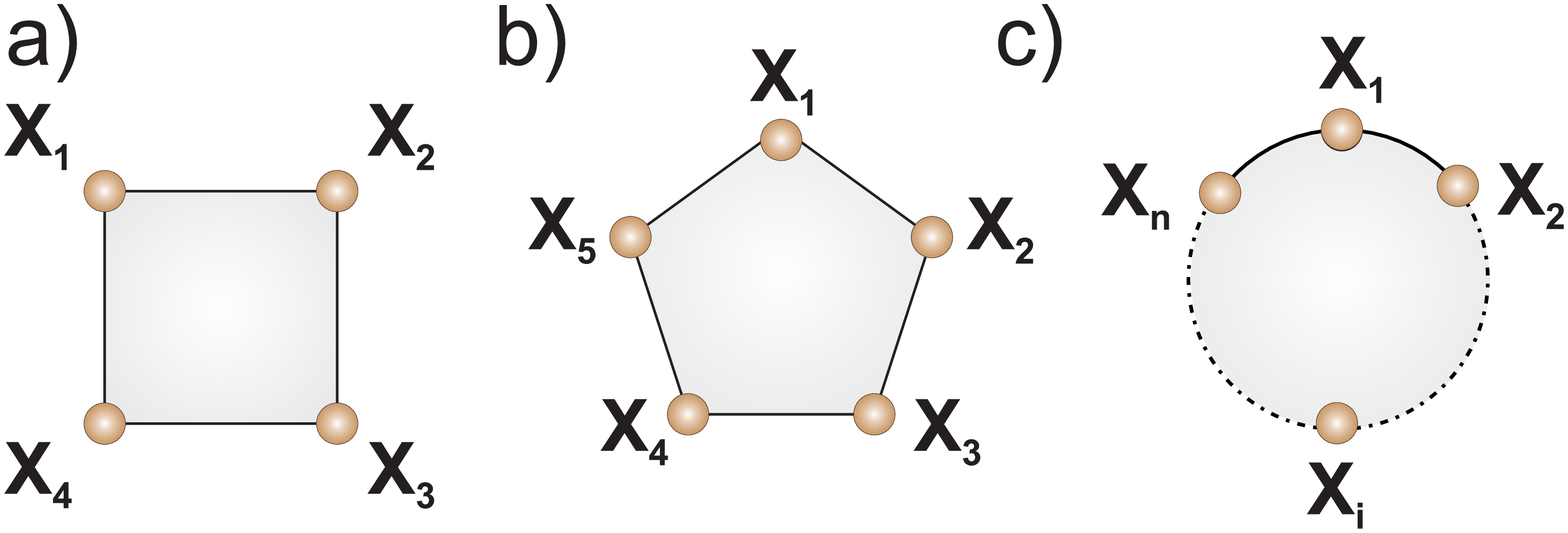}
\caption{(Color online) \label{polygons} Graphical representation of the n-cycle. The vertices represent different observables and the edges connect observables that are jointly measurable.
{\bf a.} CHSH scenario, 2 parties with 2 measurement settings each. Labelling $X_1=A_0$, $X_3=A_1$, $X_2=B_0$, $X_4=B_1$, we recover the usual picture where two parties, Alice and Bob, perform space-like separated measurements. Alice measures one out of two possible measurement settings $A_0$ or $A_1$, and similarly for Bob. {\bf b.} KCBS scenario, 5 observables arranged in a cyclic configuration such that each observable is compatible with its
neighbors {\bf c.} Generalization of the CHSH/KCBS scenario, with $n$ observables in a cyclic configuration (the $n$-cycle~\cite{Liang2011}). For dichotomic observables, the set of local/noncontextual correlations is completely characterized by (\ref{brazilian}). For a general number of outcomes, the unique nontrivial entropic inequalities are those given by~(\ref{polygon}). \label{fig:context}}
\end{center}
\end{figure}

The n-cycle scenario is defined for any number $n\geq 3$ of observables $X_1,\ldots,X_{n}$, imposing the restriction that only $X_i$ and $X_{i+1}$ are pairwise jointly measurable for all $i=1,\ldots,n$ (with $X_{n+k}=X_k$). Any two observables $X_{i}$ and $X_{i+1}$ are jointly measurable, or compatible, if the result for the measurement of $X_{i}$, even if not performed, does not depend on the prior or
simultaneous measurement of $X_{i+1}$ and vice versa. This is the notion of noncontextuality captured by the KS theorem \cite{Kochen1967}. It can readily be turned into the notion of Bell's locality  \cite{Bell1964}, where the compatibility and noncontextuality of the observables is assured by spacelike separation of local measurements of different particles. In particular, for $n=4$, the n-cycle corresponds to the CHSH Bell scenario~\cite{Clauser1969} (Fig.~\ref{fig:context}a), while for $n=5$ it is the noncontextuality scenario considered by Klyachko-Can-Binicioglu-Shumovsky (KCBS)~\cite{Klyachko2008} (Fig.~\ref{fig:context}b). For general $n$, it can be visualized as an $n$-sided polygon (Fig.~\ref{fig:context}c).

Any correlation that can be reproduced by means of a noncontextual/local hidden variable model is a convex sum of the deterministic probability distributions, that is,
\begin{equation}
p(x_{i}x_{i+1} |X_{i}X_{i+1})=\sum_{\lambda} \rho(\lambda)p(x_{i} |X_{i},\lambda) p(x_{i+1} |X_{i+1},\lambda),
\end{equation}
where $x_{i}$ stands for the outcome $x_i=0,\cdots,d-1$ of the corresponding observable $X_i$. The sum is performed over all the $d^n$ deterministic distributions parameterized by $\lambda$ (with a distribution $\rho(\lambda)$), that is, all the distributions that given a certain observable $X_{i}$ yield with probability $1$ a certain outcome $x_{i}$. The noncontextual/local deterministic points define a convex set, to which the noncontextuality/Bell inequalities are the non-trivial boundaries.

Similarly, the set of allowed nondisturbing/nonsignalling distributions also define a convex-set. Nondisturbing/nonsignalling distributions are defined as the ones for which
\begin{equation}
\begin{array}{ll}
p(x_{i}|X_{i}) & = \displaystyle \sum_{x_{i+1}} p(x_{i}x_{i+1}|X_{i}X_{i+1}) \\
& =\displaystyle \sum_{x_{i-1}} p(x_{i}x_{i-1}|X_{i}X_{i-1}),
\end{array}
\end{equation}
that is, the outcome of $X_{i}$ cannot be affected by which compatible observable $(X_{i-1}$ or $X_{i+1})$ it is jointly measured with. For dichotomic observables $x_i=0,1$ the vertices of the nondisturbing/nonsignalling set are given by \cite{Araujo2012}
\begin{equation}
p^{\gamma}_{\mathrm{max}}\left(  x_{i}x_{i+1}|X_{i}X_{i+1}\right)  =\left\{
\begin{array}{ll}
1/2 & \text{, } x_{i} \oplus x_{i+1}=\delta_{-1,\gamma_{i}}\\
0 & \text{, otherwise}%
\end{array}
\right. ,
\end{equation}
where $\oplus$ stands for addition modulo 2, $\gamma=\left\{ \gamma_{1},\ldots,\gamma_{n} \right\}$, $\gamma_{i}=\left\{-1,1 \right\}$ and the total number of $\gamma_{i}=-1$ is odd. Note that for $n=4$ and  $\gamma=\left\{ 1,1,-1,1 \right\}$ this represents the maximally nonlocal distribution allowed by nonsignalling, the so called PR-box~\cite{Popescu1994}. Given a certain nondisturbing/nonsignalling probability distribution, any other distribution than can be achieved from it using local reversible transformations is said to be equivalent. Local reversible operations consist of operations of two types: relabelling of the observables that preserve the mutual compatibility of the joint observables $X_{i}X_{i+1}$, e.g, $X_{i} \rightarrow X_{i+2}$, or relabelling of the outputs (possibly conditioned on the observable), e.g, $x_{i} \rightarrow x_{i} \oplus \delta_{i,1} $.

Recently, the full characterization of the n-cycle has been found for dichotomic observables~\cite{Araujo2012}. There are $2^{n-1}$ tight equivalent inequalities given by
\beq
\label{brazilian}
C^{\gamma}_{n}=\sum_{i} \gamma_{i} \langle X_{i}X_{i+1} \rangle  \leq  (n-2)
\eeq
where $\langle X_{i}X_{j} \rangle=p(00|X_{i}X_{j})+p(11|X_{i}X_{j})-p(01|X_{i}X_{j})-p(10|X_{i}X_{j})$ stands for the expectation value of the observable $X_{i}X_{j}$ and again the total number of $\gamma_{i}=-1$ is odd. A noncontextual/Bell inequality is said to be equivalent to another one, if it can be obtained from it by local reversible operations and/or permutation of observables. Note that $p^{\gamma}_{max}$ maximally violates $C^{\gamma}_n$, achieving $C^{\gamma}_{n}(p^{\gamma}_{max})=n$, while not violating any other equivalent inequality. In the following we will generally refer to $C_n$ and $p_{max}$ as the ones with $\gamma_{i}=1$ for $i=1,\cdots, n-1$ but $\gamma_{n}=-1$.

Similarly, the complete entropic characterization of the noncontextual/local set of probability distributions has been found for the n-cycle \cite{Fritz2012,Chaves2012} (see also \cite{Kurzy2012} for the $n=5$ case). A probability distribution in this scenario is entropically noncontextual if and only if the set of $n$ equivalent Braunstein-Caves (BC) entropic inequalities
\begin{equation}
\label{polygon}
BC^{k}_{n}=\H{{k}{k+1}} + \sum_{j\neq\, k,\,k+1} \H{j} - \sum_{j\neq k} \H{{j}{j+1}}  \leq 0
\end{equation}
hold for all $k=1,\ldots,n$ where $\H{{i}{j}}= \sum_{x_{i},x_{j}}-p(x_{i}x_{j}|X_{i}X_{j}) \log_{2} p(x_{i}x_{j}|X_{i}X_{j})$ is the Shannon entropy of the probability distribution associated with the measurements $X_{i}$ and $X_{j}$. This set of entropic inequalities is said to be maximal in the sense that no other entropic inequality can detect the contextuality/nonlocality not detectable by it, so that this set of tight entropic inequalities completely characterizes the region of noncontextual/local probabilistic models in entropy space.

In order to transform an entropic inequality into an equivalent one, the only symmetry operations that can be applied are permutations of the observables. It is a basic feature of the Shannon entropy its invariance under permutations of the sample space. Violations of BC inequalities witness then a very peculiar kind of contextuality/nonlocality. If a probabilistic distribution violates the BC inequalities, then so does any other distribution obtained by the permutation of the outcome probabilities, provided that the permutation leads to the same marginal distributions. This leads to the following phenomenon: From the point of view of the entropic inequality $BC^{k}_{n}$, the maximally contextual/nonlocal distribution $p^{\gamma}_{\mathrm{max}}$ is not different from a classically correlated and noncontextual/local distribution,
\begin{equation}
p^{\gamma^{\prime}}_{\mathrm{C}}\left(  x_{i}x_{i+1}|X_{i}X_{i+1}\right)  =\left\{
\begin{array}{ll}
1/2 & \text{, } x_{i} \oplus x_{i+1}=\delta_{-1,\gamma^{\prime}_{i}}\\
0 & \text{, otherwise}%
\end{array}
\right. ,
\end{equation}
where $\gamma^{\prime}_{i}=\gamma_{i}$ for $i \neq k$ but $\gamma^{\prime}_{k}=-\gamma_{k}$. Note that $p^{\gamma^{\prime}}_{\mathrm{C}}$ has an even total number of $\gamma^{\prime}_{i}=-1$ and does not violate any $C_{n}^{\gamma}$ inequality. Thus, since $p^{\gamma}_{\mathrm{max}}$ is entropically equivalent to a noncontextual/local distribution, it does not violate any entropic inequality. In this sense, an entropic noncontextual/Bell inequality is a necessary but not sufficient criterion to probe the noncontextual/local behavior a distribution. However, as we show next, entropic inequalities can be turned into a necessary and sufficient condition, since any contextual/nonlocal distribution violating $C_n^{\gamma}$ also violates the entropic BC inequalities when properly mixed with a noncontextual/local distribution.

\section{Entropic inequalities completely characterize the n-cycle scenario with dichotomic outcomes}
\label{sec:equivalence}

First, note that the maximum violation of the BC inequality for dichotomic observables is given by $BC^{k}_{n}=1$, since $\H{j} \leq \H{{j}{j+1}}$ and $\H{{k}{k+1}} \leq \H{k} + H(X_{k+1}) \leq H(X_{k + n-1}X_{k}) + H(X_{k+1}) \leq H(X_{k + n-1}X_{k})+ 1 $. The maximal violation of $BC^{k}_{n}=1$ can be achieved by the probability distribution
\begin{equation}
\label{Emax}
p^{\gamma}_{Emax}=\frac{1}{2}(p^{\gamma}_{\mathrm{max}}+p^{\gamma^{\prime}}_{\mathrm{C}}),
\end{equation}
for all $\gamma$ such that $\gamma_{k}=-1$, since $\H{j} = \H{{j}{j+1}}=1$ for all $j \neq k$ and $\H{{k}{k+1}}=2$. That is, a convex combination of two nonviolating distributions may violate an entropic inequality, highlighting the strongly non-linear character of it. Note that $p^{\gamma^{\prime}}_{\mathrm{C}}$ can be achieved with 1 bit of shared randomness, that is, $x_{i}=x_{i+1}=0$ or $x_{i}=x_{i+1}=1$ with the same probability 1/2 for all $i=1,\ldots,n$. Similarly the convex combination in (\ref{Emax}) also requires 1 bit of shared randomness, outputting $p^{\gamma}_{\mathrm{max}}$ or $p^{\gamma^{\prime}}_{\mathrm{C}}$, both with probability 1/2. When augmented with some shared randomness, it is possible to turn an entropically noncontextual/local distribution into a contextual/nonlocal one. In a noncontextual/Bell scenario, shared randomness is always an available and valid resource. However, for usual noncontextual/Bell inequalities, the mixing with a local point cannot improve the violation of the inequality due to its linearity.

As we show next, the mixing with $p^{\gamma^{\prime}}_{\mathrm{C}}$ is sufficient to entropically detect any contextual/nonlocal distribution. First we note that in the CHSH scenario ($n=4$ in the n-cycle), it is known that any probability distribution can be transformed into an isotropic distribution through a local depolarization process, keeping the $C_{4}$ value invariant~\cite{Masanes2006}. The isotropic distribution has the property of being invariant under the interchange of the inputs or outputs and being locally unbiased, that is, $p_{\mathrm{I}}(x_{i}x_{i+1}|X_{i}X_{i+1})=p_{\mathrm{I}}(x_{i+1}x_{i}|X_{i}X_{i+1})$ and $p_{\mathrm{I}}(x_{i}|X_{i})=1/2$.  For the n-cycle, the isotropic distribution $p_{\mathrm{I}}$ corresponds to a probabilistic mixture,
\begin{equation}
p_{\mathrm{I}}=\epsilon p_{\mathrm{max}}+(1-\epsilon)p_{\mathrm{w}},
\end{equation}
with $p_{\mathrm{w}}(x_{i}x_{i+1}|X_{i}X_{i+1})=1/4$ being pure white noise. The corresponding  $C_n$ value is given by $C_{n}(p_{\mathrm{I}}) = n\epsilon$, that is, $p_{\mathrm{I}}$ is contextual/nonlocal for $\epsilon > (n-2)/n$. From the entropic point of view $p_{\mathrm{I}}$ is equivalent to the distribution $\epsilon p_{\mathrm{C}}+(1-\epsilon)p_{\mathrm{w}}$ and thus no direct violation of entropic BC inequalities is possible.

The first step in our proof is to show that any distribution in the n-cycle scenario can be turned into a isotropic one without changing the values of $C_n$, that is, a generalization of the depolarization protocol devised in Ref. ~\cite{Masanes2006} (see also~\cite{Grudka2012}) for the $n=4$ case. Then, for our purposes and to simplify the presentation it is enough to consider isotropic boxes only (however as shown in the Appendix this is not strictly necessary). The depolarization procedure is done in two steps:

i) $p(x_{i}x_{j}|X_{i}X_{j})$ is made locally unbiased by flipping both outputs simultaneously with probability 1/2, that is, $x_{i} \rightarrow x_{i} \oplus 1$ and $x_{j} \rightarrow x_{j} \oplus 1$

ii)$p(x_{i}x_{j}|X_{i}X_{j}) \rightarrow \displaystyle \sum_{k} p(\bar{x}^{k}_{i}\bar{x}^{k}_{j}|X_{i\oplus_{n}k}X_{j\oplus_{n}k} )$ where $\bar{x}^{k}_{i}$ means flipping the output if $i \in \left\{ 1, \ldots ,n-k+1 \right\}$ and $\oplus_{n}$ stands for addition modulo $n$. After the second step the initial distribution is in the isotropic form, however maintaining the value of $C_n$ unchanged.

Computing the BC value for the isotropic distribution mixed with the classically correlated box, that is, $v p_{\mathrm{I}}+ (1-v)p_{\mathrm{C}}$, expanding around $v=0$ we find that
\begin{equation}
BC_{n}=\frac{v}{\ln{4}} \left[ f(n,\epsilon)-(2- n(1-\epsilon) ) \ln v \right]
\end{equation}
with $f(n,\epsilon)= 2 - n(1-\epsilon)(1+\ln{2}) + (n+\epsilon -n\epsilon)\ln{(1-\epsilon)} -\epsilon\ln{(1+\epsilon)} + \ln{(4/(1-\epsilon^{2}))} $. For any $2- n(1-\epsilon)>0$ taking a sufficiently small $v$ ensures BC to be positive since $f(n,\epsilon)$ does not depend on $v$. That is, for any nonlocal isotropic distribution $\epsilon > (n-2)/n$ the BC inequality can be violated. This is the same bound obtained by the direct calculation of $C_n(p_{\mathrm{I}})$. Given that any distribution can be turned into the isotropic box without changing its $C_n$ value, this means that \emph{any contextual/nonlocal distribution in the n-cycle scenario also violates the entropic $BC_{n}$ inequality when properly mixed with a classically correlated and noncontextual/local distribution}.

We note that we have analyzed the specific case of a distribution violating a specific equivalence of $C^{\gamma}_n$ where all $\gamma_{i}=1$ but $\gamma_{n}=-1$. However, the generalization for distributions violating other equivalences of $C^{\gamma}_{n}$ is straightforward. Since there is a one-to-one correspondence between the vertices of the nondisturbing/nonsignalling set and each of the facets of the noncontextual/local set, if a distribution violates $C^{\gamma}_{n}$, it can be brought to the isotropic form $p^{\gamma}_{\mathrm{I}}=p^{\gamma}_{\mathrm{max}}+p_{\mathrm{w}}$ again without changing the value of $C^{\gamma}_{n}$. Mixing $p^{\gamma}_{\mathrm{I}}$ with the corresponding classical correlation $p^{\gamma^{\prime}}_{\mathrm{C}}$ will lead to violations of $BC^{k}_{n}$ and therefore to the same conclusions as before. It is important to note that the characterization of the noncontextual/local set is achieved with $n$ entropic inequalities, as opposed to the $2^{n-1}$ linear inequalities (\ref{brazilian}), however at the cost of introducing a list of classical points $p^{\gamma^{\prime}}_{\mathrm{C}}$.

\section{Conclusion}
\label{sec:conclusion}
In principle, entropic inequalities only provide a necessary but not sufficient criterion for noncontextuality and local realism. However, we have shown that for the n-cycle with dichotomic outcomes, entropic inequalities turn also to be sufficient, since any contextual/nonlocal probabilistic model will display entropic violations if properly mixed with a classical model. It is quite surprising that a polynomial number of non-linear and non-tight inequalities may completely characterize the set of noncontextual/local correlations, that otherwise would require an exponential number of linear and tight inequalities to do so.

One obvious question is how this result would extend for more complex scenarios, involving more than two-outcomes and possibly more parties as in a multipartite Bell test. Even in the bipartite case, $n=4$ for the n-cycle, the complete characterization of the local correlations is not known for a number of outcomes larger than 3 \cite{Masanes2002,Bancal2010}. Could it be that the BC entropic inequality augmented with shared randomness fully characterizes the n-cycle for a general number of outputs? Another interesting question is to understand the role of entropic inequalities in the bilocality scenario, where independence constraints define a nonconvex set, difficult to characterize in the probability space \cite{Branciard2010}.

Finally, violations of linear Bell inequalities can be understood as a resource, for instance, allowing for higher probability of success in some information tasks \cite{Buhrman2010}. Is there any operational interpretation for the violation of an entropic inequality in terms of a relevant physical task? If that turns out to be the case, an interesting scenario would arise, where nonlocal but entropically classical correlations could be turned into a useful resource, being activated by the use of shared randomness.

\appendix*
\section{Violation of the entropic inequalities without the depolarization procedure}
We show here that given a general nonlocal probability distribution in the CHSH scenario, the mixing with the classical correlation is sufficient to violate the entropic inequality, without the need of the depolarization procedure.

The local set consists of 16 extremal points $p^{\alpha,\beta,\gamma,\delta}_{det}$ parameterized as
\begin{equation}
p_{det}\left(  ab|xy\right) = \left\{ \begin{array}{r@{\quad}c@{\quad}l} 1 &,&
a=\alpha x \oplus \beta,\\
&& b=\gamma y \oplus \delta\\
0 &,& \mbox{otherwise}
\end{array} \right. ,
\end{equation}
and all the 8 nonlocal extremal points $p^{\alpha,\beta,\gamma}_{PR}$ of the nonsignalling set can be parameterized as
\begin{equation}
p_{PR}\left(  ab|xy\right)  =\left\{
\begin{array}{ll}
1/2 & \text{, } a\oplus b=xy+\alpha x +\beta y +\gamma\\
0 & \text{, otherwise}%
\end{array}
\right. ,
\end{equation}
here $\alpha,\, \beta,\,\gamma,\delta \in \{0,1\}$. To simplify the description we have employed the common notation to the CHSH scenario, that is, $X_1$ and $X_3$ corresponding to $x=0$ and $x=1$ and $X_2$ and $X_4$ corresponding to $y=0$ and $y=1$, while $a$ and $b$ label the corresponding outcomes. For a general distribution written as a convex combination of all extreme points
\begin{equation}
p\left(  ab|xy\right)= \sum_{\alpha,\beta,\gamma} \varrho^{\alpha,\beta,\gamma}_{PR} p^{\alpha,\beta,\gamma}_{PR} + \sum_{\alpha,\beta,\gamma,\delta} \varrho^{\alpha,\beta,\gamma,\delta}_{det} p^{\alpha,\beta,\gamma,\delta}_{det},
\end{equation}
the condition for the violation of the CHSH inequality is that $C_4= \displaystyle \sum_{\alpha,\beta,\gamma,\delta} (-1)^{(1-\alpha)(\beta+\delta)+\alpha(\beta+\gamma+\delta)} \varrho^{\alpha,\beta,\gamma,\delta}_{det} + 2\varrho^{0,0,0}_{PR} - 2 \varrho^{0,0,1}_{PR} >1$. Mixing the $p\left(  ab|xy\right)$ with the classical correlated box as before and expanding around $v=0$, one finds that $BC_4=(v/\ln{4})\left[ g+2\ln{v}(1-C_4) \right]$, that violates the entropic inequality for $C_4 > 1$ ($g$ is a function of $p\left(  ab|xy\right)$ but independent of $v$).

The same argument can be applied to the general n-cycle, and we have tested up to $n=7$ that the same result holds. For larger n, the procedure becomes unfeasible given that the number of extremal points increases exponentially, but we conjecture that the same result should hold for any n.

\begin{acknowledgments}
I would like to thank D. Gross for helpful discussions and J. B. Brask, T. Fritz and M. T. Quintino for comments on the manuscript. This work was supported by the Excellence Initiative of the German Federal and State Governments (Grant ZUK 43).
\end{acknowledgments}

\bibliography{ncyclebib}

\end{document}